# Realization of Dirac Point with Double Cones in Optics

*Li-Gang Wang*[1,2], *Zhi-Guo Wang*[1,3], *and Shi-Yao Zhu*[1,2,4]

[1]*Center of Optical Sciences and Department of Physics, The Chinese University of Hong Kong, Shatin, N.T., Hong Kong*
[2]*Department of Physics, Zhejiang University, Hangzhou 310027, China*
[3]*Department of Physics, Tongji University, Shanghai 200092, China*
[4]*Department of Physics, Hong Kong Baptist University, Kowloon Tong, Hong Kong*

## Abstract

The Dirac point with a double-cone structure for optical fields, an optical analogy Dirac fermions in graphene, can be realized in optically homogenous metamaterials. The condition for the realization of Dirac point in optical systems is the varying of refractive index from negative to zero and then to positive. Our analytical and numerical analysis have verified that, similar to electrons in graphene, the light field near the Dirac point possesses of the *pseudodiffusive* property, obeying the $1/L$ scaling law, where $L$ is the propagating distance of light inside the media.

Recently the discovery of graphene [1-2] refreshes attention to the two-dimensional massless Dirac equation. In graphene (a single atomic layer of carbon atoms), the conduction band and the valence band touch each other at frequency $\omega_D$ forming the Dirac point with a double-cone structure. Near the Dirac point, the dispersion is linear with two branches. Due to these facts, graphene is not only interesting because of its electronic properties, such as the frequency dependent conductivity [3-7], collective excitations (plasmons) [8-11], and thermoplasma polaritons [12], but also being an excellent candidate for experimental demonstration of the quantum relativistic effect, such as *Zitterbewegung* effect [13-15] and Klein paradox [16], which were predicted in quantum electrodynamics for relativistic electrons in vacuum but never observed directly.

In this Letter, we perform a theoretical investigation on realizing the Dirac point with a double-cone structure in homogenous materials for light fields, thus provide a direct optical analog of graphene. It is a known fact that, compare to solids, optical systems offer clean and easy controlled way to experimentally test theoretical predictions. The experimental test in electronic systems is usually hindered by the difficulty to maintain system homogeneity. While for optical systems, no such difficulty exists. So establishing the optical analog of graphene would open up the possibility to study condensed matter analogies in optical way. In optics, the Maxwell's equations for electromagnetic waves can be reduced to the Helmholtz equation. For a homogenous medium, when a light field is polarized in the $z$ direction, the Helmholtz equation could be written as $\left(\frac{\partial}{\partial x} + i\frac{\partial}{\partial y}\right)\left(\frac{\partial}{\partial x} - i\frac{\partial}{\partial y}\right) E_z(x,y,\omega) = (ik(\omega))^2 E_z(x,y,\omega)$, with a wave number $k(\omega)$. This equation can be written in the form of the Dirac equation

$$\begin{bmatrix} 0 & -i\left(\frac{\partial}{\partial x} - i\frac{\partial}{\partial y}\right) \\ -i\left(\frac{\partial}{\partial x} + i\frac{\partial}{\partial y}\right) & 0 \end{bmatrix} \begin{pmatrix} E_{z1}(x,y,\omega) \\ E_{z2}(x,y,\omega) \end{pmatrix} = k(\omega) \begin{pmatrix} E_{z1}(x,y,\omega) \\ E_{z2}(x,y,\omega) \end{pmatrix}, \quad (1)$$

where $E_{z1}(x,y,\omega)$, $E_{z2}(x,y,\omega)$ are two eigenfunctions of the electrical fields corresponding to the same $k(\omega)$ [17]. In most conventional optical media or free space $k(\omega)$ is always positive-definite because of causality. That is to say, it seems one may not have the Dirac point with a double-cone structure as in the electronic systems. However, this should not prevent us from realizing Dirac point in some particular optical systems. For example, from the similarity of the photonic bands in the two-dimensional photonic crystal (2DPC) with the electron bands in solids, it was noticed that the band gap of the 2DPC on a triangular or honeycomb lattice could become vanishingly small at corners of the Brillouin zone to form the Dirac point with double-cone structure [18-20]. Several comparisons between the electronic and optical transport properties near the Dirac point were made [20-22], such as the extremal transmission possessing a new

"pseudodiffusive" scaling [20] and the demonstration of the photon's *Zitterbewegung* effect [22] near the Dirac point in the 2DPC. Naturally, it would be of great interest to find out what is the condition to have the Dirac dispersion for the light field in a homogenous medium.

Generally, the wave vector of a medium, $k(\omega)$, can be expanded as $k(\omega) = k(\omega_D) + (\omega - \omega_D)/v_D + \beta(\omega - \omega_D)^2 + \cdots$ at frequency $\omega_D > 0$, with the group velocity $v_D = (d\omega/dk)\big|_{\omega=\omega_D}$. If $k(\omega_D) = 0$ and the quadratic- and higher-order terms could be neglected, then we would have a linear dispersion

$$k(\omega) = (\omega - \omega_D)/v_D. \qquad (2)$$

For media satisfying Eq. (2) with $\omega_D > 0$, the wave number, $k(\omega)$, varies from negative to zero and then to positive with the frequency, and so does the refractive index of the medium with zero refractive index at $\omega_D$. We call such media the negative-zero-positive index (NZPI) metamaterials. Substituting Eq. (2) into Eq. (1), we have

$$\begin{bmatrix} 0 & -iv_D\left(\dfrac{\partial}{\partial x} - i\dfrac{\partial}{\partial y}\right) \\ -iv_D\left(\dfrac{\partial}{\partial x} + i\dfrac{\partial}{\partial y}\right) & 0 \end{bmatrix} \begin{pmatrix} E_{z1}(x,y,\omega) \\ E_{z2}(x,y,\omega) \end{pmatrix} = (\omega - \omega_D)\begin{pmatrix} E_{z1}(x,y,\omega) \\ E_{z2}(x,y,\omega) \end{pmatrix}. \qquad (3)$$

Note that the positive and negative branches of the band structure coexist (see Fig. 1(b)). Equation (3) is the massless Dirac equation of the light fields in homogenous materials, which is the same as that of electrons in graphene [23-25]. Therefore, for the NZPI metamaterial [satisfying Eq. (2)], we will have the Dirac point with a double-cone structure for the light field at frequency $\omega_D$. It is expected that the propagation of the light field is analogous to that of the electron in graphene. In fact, in the 2DPC [18], the effective refractive index varies from negative to positive near this Dirac point. At $\omega_D$, Eq. (3) takes the form of diffusion equation,

$$-iv_D\left(\frac{\partial}{\partial x}+i\frac{\partial}{\partial y}\right)E_{z1}(x,y,\omega)=0, \tag{4a}$$

$$-iv_D\left(\frac{\partial}{\partial x}-i\frac{\partial}{\partial y}\right)E_{z2}(x,y,\omega)=0, \tag{4b}$$

which are the same as that of the massless Dirac equation at zero energy. [23] Therefore, it can be predicted that the behavior of light fields at $\omega_D$ has the diffusive properties inside the medium of Eq. (2), like electrons at Dirac point of graphene [2, 23-25]. At or near $\omega_D$, because of $k^2=k_x^2+k_y^2\to 0$, the $k_x$ component becomes a pure imaginary number (i.e., $k_x=i|k_y|$) for any real $k_y$, thus the fields along the $x$ direction between the interval $L$ have the relation

$$t(L,k_y)=\frac{E(L)}{E(0)}=\exp(-|k_y|L), \tag{5}$$

then the total energy transmittance is

$$T_{All}=\int_{-\infty}^{\infty}|t(L,k_y)|^2 dk_y=1/L, \tag{6}$$

which tells us that the propagation of light field at (or near) $\omega_D$ exhibits the $1/L$ scaling, a main characteristic of the diffusion phenomenon.

Now the question is how to find such a medium satisfying Eq. (2) with $\omega_D>0$. Fortunately, the man-made metamaterials with small absorption [26-30] can meet the requirement. The NZPI metamaterial was reported in Ref. [27]. The permittivity and permeability of the metamaterials are determined, respectively, by $\varepsilon_1(\omega)=1-\omega_{ep}^2/(\omega^2+i\gamma_e\omega)$ and $\mu_1(\omega)=1-\omega_{mp}^2/(\omega^2+i\gamma_m\omega)$ [28], where $\omega_{ep}$ and $\omega_{mp}$ are, respectively, the electronic plasma frequency and the magnetic plasma frequency, and $\gamma_e$ and $\gamma_m$ are the damping rates relating to the absorption of the material. We assume $\gamma_e=\gamma_m=\gamma\ll\omega_{ep}$, $\omega_{mp}$. Several recent developments [26-27, 29-30] have shown that $\omega_{ep}$ and $\omega_{mp}$ are artificially adjustable by changing the metamateral structure in GHz region. Obviously when $\omega_{ep}=\omega_{mp}=\omega_D$ and $\gamma=0$ (no loss), both $\varepsilon_1(\omega_D)$ and $\mu_1(\omega_D)$ are zero, and it is easy to show that Eq. (2) is valid near

$\omega_D$.

To illustrate the properties of the light propagation near the Dirac point, we first consider a semi-infinite homogenous NZPI metamaterial, as shown in Fig. 1(a). Such set up reduces the non-ideal coupling effect of light fields at the interface between the NZPI metamaterial and the outside space and demonstrates the physics of light fields near the Dirac point. The light field is incident from the vacuum to the surface ($x=0$) and propagates to $x=L$. The transmission coefficient at position ($x=L$) is given by [31] $t(k_y,\omega) = \alpha \exp[ik_{x1}L]$, where $\alpha = \dfrac{2(q_0 q_m)^{1/2}}{(q_0 + q_m)}$ is determined by the boundary condition, $k_{x1} = \sqrt{\varepsilon_1}\sqrt{\mu_1}\sqrt{k_0^2 - \dfrac{k_y^2}{\varepsilon_1 \mu_1}}$ is the $x$ component of the wave number inside the semi-infinite metamaterial for $\varepsilon_1 \mu_1 k_0^2 > k_y^2$ and otherwise $k_{x1} = i\sqrt{\varepsilon_1}\sqrt{\mu_1}\sqrt{\dfrac{k_y^2}{\varepsilon_1 \mu_1} - k_0^2}$, $q_m = \dfrac{k_{x1}}{\mu_1 k_0}$, $k_y$ is the $y$ component of the wave number, $k_0 = \dfrac{\omega}{c}$ is the wave number in the vacuum, and $q_0 = \sqrt{k_0^2 - k_y^2}/k_0$ for $k_0 > k_y$ and otherwise $q_0 = i\sqrt{k_y^2 - k_0^2}/k_0$. Near $\omega_D$, the transmission coefficient at $x=L$ becomes $t(k_y,\omega) = \alpha \exp[-|k_y|L]$. Assume $\alpha$ be independent of $k_y$ (the ideal interface), then the total transmittance is calculated by $T_{All} = \int_{-\infty}^{\infty} |t(k_y,\omega)|^2 dk_y = |\alpha|^2/L$ (different from Eq. (6) by the coupling coefficient due to the incident surface). Regardless of the boundary condition at $x=0$, the light transport at and near $\omega_D$ is inversely proportional to the propagating distance $L$ inside the homogeneous NZPI metamaterial (the $1/L$ scaling).

In Fig. 1(b), we plot the transmitted spectrum of light after propagating a distance $L = 1000\,\text{mm}$ inside the semi-infinite metamaterial. It is clearly seen that both the upper and lower passbands touch at $\omega_D/2\pi = 10\,\text{GHz}$ (a small touch region) and nearby the dispersion is linear. It is expected [see Fig. 1(b)] that as $L$ becomes larger, the transition from the higher transmittance (passbands) to the lower transmittance (gap regions) will become smaller, and the touch at $\omega_D$ will become an ideal point.

The spatial-evolution of the total electric field for a normally incident Gaussian beam is plotted in Figs. 2(a)-2(c) for different frequencies. The arrows in $x > 0$ denote the relative magnitude and direction of the energy flow $\vec{S}$ (Poynting vector). Here the angular spectral amplitude of the incident Gaussian beam is given by $E(k_y, 0) = \frac{W}{\sqrt{2}} \exp[-W^2 k_y^2 / 4]$ with a beam half-width $W$. Within the upper ($\omega/2\pi = 13$ GHz) and lower passbands ($8$ GHz), we have light propagation effects, see Figs. 2(a) and 2(c). At the Dirac point, $\omega/2\pi = 10$ GHz, the evolution of the light field, shown in Fig. 2(b), becomes a static diffusive radiation.

In order to observe the $1/L$ scaling behavior [Eq. (6)] near the Dirac point, we define a characteristic quantity $\xi = S_r \times L$ to describe the light transport inside the medium, where $S_r \equiv S(x \geq 0, y = 0)/S_0$ is the relative energy flow along the $x$ axis, and $S_0 \equiv S(x = 0, y = 0)$, which depends on the coupling strength. The dependence of the quantity $\xi$ on the propagating distance $L$ under different frequencies is plotted in Fig.3, which shows that near the Dirac point ($\omega/2\pi = 10$, 10.2GHz) the value of $\xi$ tends to be a constant with the increase of $L$. Therefore, the light transports $S_r$ near the Dirac point obey the $1/L$ scaling law. Away from the Dirac point ($\omega$ moving into the passband), $\xi$ gradually becomes linear dependence of $L$, which means almost a constant $S_r$ except small absorption. For a comparison, the dotted line denotes the light frequency is in the completely transparent passband.

Realistically the semi-infinite structure does not exist in nature. So next we consider the propagation of light through a homogenous slab system. Figure 4 shows the change of $\xi$ as a function of $x$ inside the slabs with different thicknesses $d$. As $d$ increases, the change of $\xi$ inside the slab approaches to the limit of $d \to \infty$, as discuss above. For a finite $d$, $\xi$ always initially increases and then gradually decays in order to match the second boundary condition at $x = d$, which is indicated by crossed points in

Fig. 4. From Fig. 4, we conclude that for a sufficient thick slab, the light energy transport obeys the $1/L$ scaling law.

In the above calculations the absorption is relatively low. Large absorption will destroy the $1/L$ scaling law. For low absorption, the coupling strength is low near the Dirac point [see the small values in Fig. 2(b) and in the inset of Fig. 3]. How to enhance the coupling strength of the light field at interface near the Dirac point should be further investigated for its potential applications.

In a summary, it is the first time, to the best of our knowledge, that the condition for the realization of Dirac point physics in optical systems is presented. We show that the Dirac point with the double-cone structure exists in homogeneous medium as long as Eq. (2) is valid. The existing NZPI metamaterials is a good candidate. Our analytical and numerical analysis have verified that the light field near the Dirac point becomes a diffusive wave and its energy transport obeys the $1/L$ scaling law inside the NZPI medium, manifests the behavior of electrons in graphene.

L. G. Wang would like to thank Prof. H. Q. Lin for his helpful discussions. This work is supported by CUHK 401806 and HKUST 3/06C of HK Government, FRG of HKBU, and NSFC (Contract No. 10604047).

# Reference

[1] K.S. Novoselov, et al., Science **306**, 666 (2004).

[2] K.S. Novoselov, et al., Nature (London) **438**, 197 (2005).

[3] V. P. Gusynin and S. G. Sharapov, Phys. Rev. B **73**, 245411(2006).

[4] L. A. Falkovsky and A. A. Varlamov, Eur. Phys. J. B **56**, 281 (2007).

[5] V. P. Gusynin, S. G. Sharapov, and J. P. Carbotte, Phys. Rev. Lett. **96**, 256802 (2006).

[6] A.B. Kuzmenko, E. van Heumen, F. Carbone, and D. van der Marel, Phys. Rev. Lett. **100,** 117401 (2008).

[7] G.W. Hanson, J. Appl. Phys. **103**, 064302(2008).

[8] V. Ryzhii, Jpn. J. Appl. Phys. **45**, L923 (2006).

[9] V. Ryzhii, A. Satou, and T. Otsuji, J. Appl. Phys. **101**, 024509 (2007).

[10] V. Apalkov, X.-F. Wang, and T. Chakraborty, International J.Mod. Phys. B **21**, 1165 (2007).

[11] E. H. Hwang and S. Das Sarma, Phys. Rev. B **75**, 205418 (2007).

[12] O. Vafek, Phys. Rev. Lett. **97**, 266406 (2006).

[13] E. Schrodinger, Sitzungsber. Preuss. Akad. Wiss., Phys. Math. Kl. **24**, 418 (1930).

[14] H. Feschbach and F. Villars, Rev. Mod. Phys. **30**, 24 (1958).

[15] József Cserti and Gyula Dávid, Phys. Rev. B **74**, 172305 (2006).

[16] O. Klein, Z. Phys. **53**, 157 (1929).

[17] P. A. M. Dirac, The Principles of Quantum Mechanics, 4th edn. Oxford U. P., (1958)

[18] M. Notomi, Phys. Rev. B **62**, 10696 (2000).

[19] F. D. M. Haldane and S. Raghu, Phys. Rev. Lett. **100**, 013904 (2008).

[20] R. A. Sepkhanov, et al., Phys. Rev. A **75**, 063813 (2007).

[21] R. A. Sepkhanov, et al., Phys. Rev. B **78**, 045122 (2008).

[22] X. Zhang, Phys. Rev. Lett. **100**, 113903 (2008).

[23] M. I. Katsnelson, Eur. Phys. J. B **51**, 157 (2006).

[24] J. Tworzydlo, et al., Phys. Rev. Lett. **96**, 246802 (2006).

[25] K. Nomura and A. H. MacDonald, Phys. Rev. Lett **98**, 076602 (2007).

[26] D. R. Smith, D. R. Smith, S. Schultz, Science **292**, 77 (2001).

[27] F. Zhang, et al., J. Appl. Phys. **103**, 084312 (2008).

[28] R. W. Ziolkowski, Phys. Rev. E **70**, 046608 (2004).

[29] J. B. Pendry, et al., Phys. Rev. Lett. **76**, 4773 (1996).

[30] J. B. Pendry, et al., IEEE Trans. Microwave Theory Tech. **47**, 2075 (1999).

[31] M. Born and E. Wolf, *Principles of Optics*, (Cambridge University Press, Cambridge, 1999).

# FIGURE CAPTIONS

FIG. 1. (Color Online). (a) Schematic of a semi-infinite NZPI metamaterial and (b) transmitted spectral distribution at position $L$. Red dashed lines denote the range of the light cone, and blue and white areas denote the high transmission (passbands) and the prohibition of light (gaps), respectively. The parameters are $\omega_{ep} = \omega_{mp} = \omega_D = 2\pi \times 10$ GHz and $\gamma = 10^{-5}$ GHz.

FIG. 2. Evolutions of the total electric fields for a narrowed Gaussian beam through the interface at different frequencies: $\omega/2\pi = 13$ GHz (a), 10GHz (b) and 8GHz (c), with $W = 5\lambda_0$, here $\lambda_0$ corresponds to $\omega_D$. Other parameters are the same as in Fig. 1 except for $\gamma = 10^{-4}$ GHz.

FIG. 3. Change of $\xi$ as a function of $L$ inside the semi-infinite structures under different frequencies. Inset shows the change of $S_0$ with frequency. Other parameters are the same as in Fig. 2.

FIG. 4. Change of $\xi$ as a function of the propagating position $x$ inside the different finite slabs, with $\omega = \omega_D$ and $W = 10\lambda_0$. Other parameters are the same as in Fig. 2.

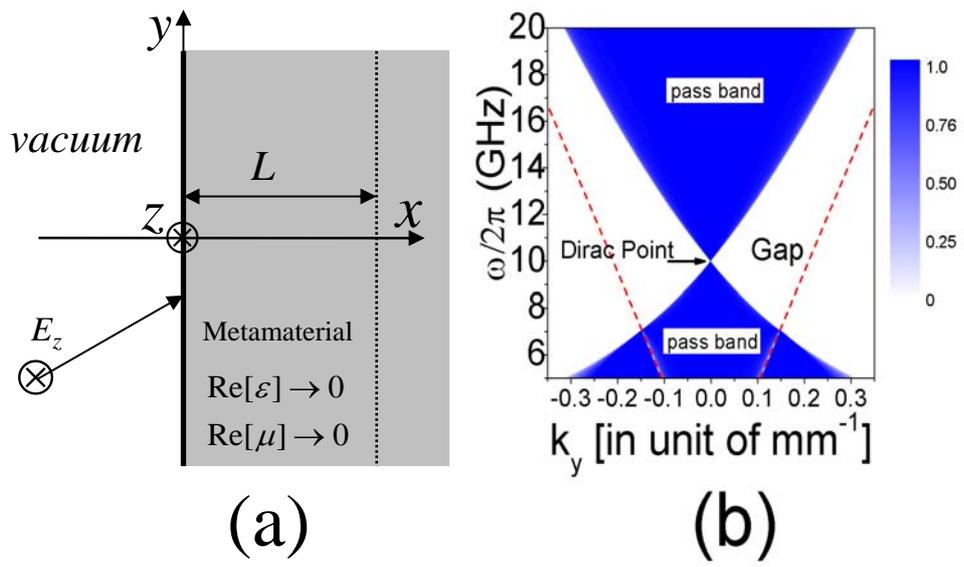

**FIG. 1**

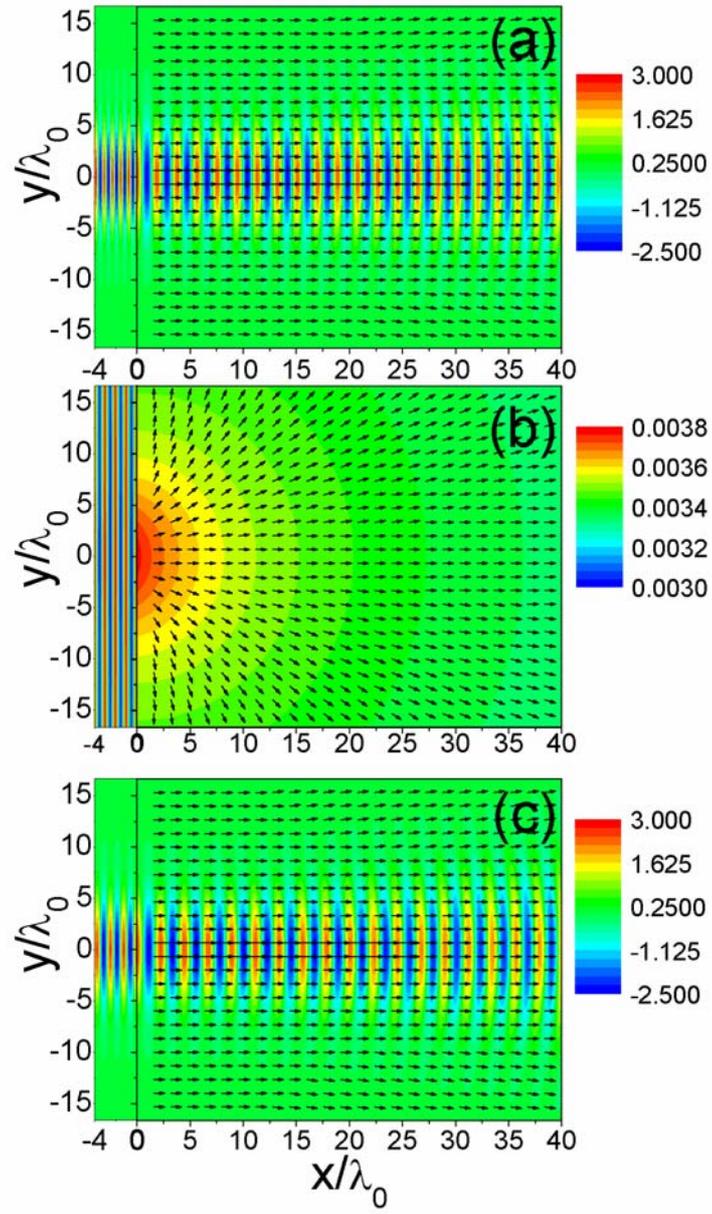

**FIG. 2**

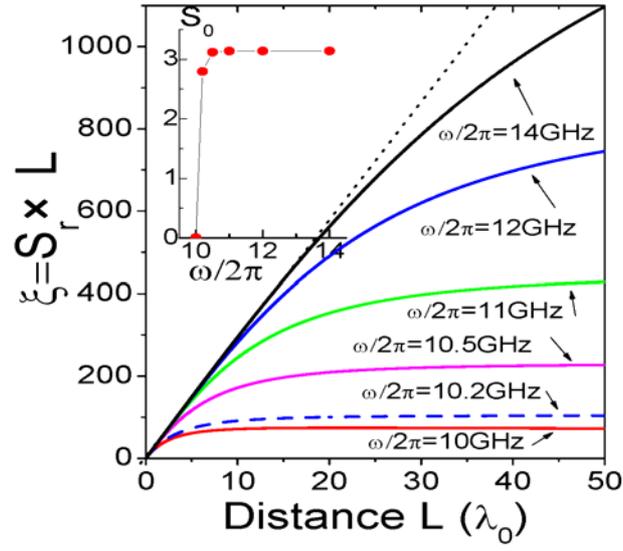

**FIG. 3**

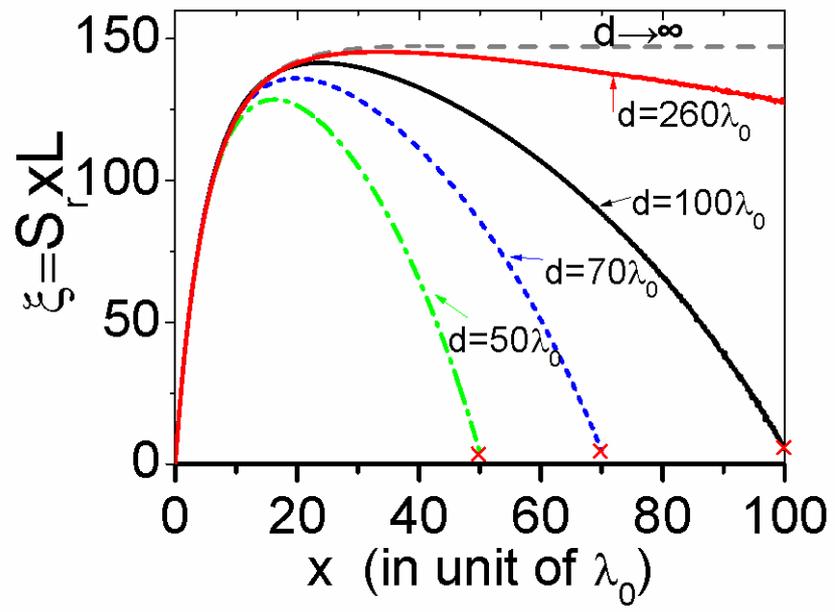

**FIG. 4**